\documentclass[a4paper, amsfonts, amssymb, amsmath, preprint, showkeys, nofootinbib]{revtex4-1}
\usepackage[english]{babel}
\usepackage[utf8]{inputenc}
\usepackage[colorinlistoftodos, color=green!40, prependcaption]{todonotes}

\usepackage[pdftex, pdftitle={Article}, pdfauthor={Author}]{hyperref} % For hyperlinks in the PDF
\begin{document}

\title{Generalized free energy landscape of black hole phase transition}

\author{Ran Li$^{a,b}$}
\email{liran@htu.edu.cn}
    
\author{Jin Wang$^{b,c}$}
\email{Corresponding author: jin.wang.1@stonybrook.edu}

\affiliation{$^a$ School of Physics, Henan Normal University, Xinxiang 453007, China\\
$^b$ Department of Chemistry, Stony Brook University, Stony Brook, NY 11794, USA \\
$^c$ Department of Physics and Astronomy, Stony Brook University, Stony Brook, NY 11794, USA}

\date{\today} % Leave empty to omit a date

\begin{abstract}
Recently, the stochastic dynamical model based on the free energy landscape was proposed to quantify the kinetics of the black hole phase transition. An essential concept is the generalized free energy of the fluctuating black hole, which was defined in terms of the thermodynamic relation previously. In this work, by employing the Gibbons-Hawking path integral approach to black hole thermodynamics, we show that the generalized free energy can be derived from the Einstein-Hilbert action of the Euclidean gravitational instanton with the conical singularity. This work provides a concrete and solid foundation for the free energy landscape formalism of black hole phase transition.
\end{abstract}


\maketitle

%\tableofcontents

\section{Introduction}
\label{sec:intro}

Recently, it was proposed that the kinetics of black hole phase transition can be described by the stochastic dynamical model based on the free energy landscape \cite{Li:2020khm,Li:2020nsy,Li:2021vdp}. An essential concept is the generalized free energy function of the fluctuating black hole, which was defined in terms of the thermodynamic relation previously. It is well known that thermodynamics is universal. Therefore, the previous definition of the generalized free energy from the thermodynamics viewpoint should be reasonable. However, how to derive the generalized free energy of the fluctuating black hole from the first principle is still unclear. In the gravity aspect, this thermodynamic quantity should be related to the Einstein-Hilbert action in the path integral formalism of quantum gravity \cite{Gibbons:1976ue}. In this paper, we address this issue. Prior to this, we will make some clarifications.

In studying phase transition, order parameter is usually used to distinguish two different phases (or orders). For liquid-gas transitions, the order parameter is the density which has a clear distinction between the liquid phase and the gas phase. For the Reissner-Nordstrom Anti-de Sitter (RNAdS) black hole phase transition in the extended phase space, the number density of the black hole molecules is proposed to distinguish the different black hole phases \cite{Wei:2015iwa,Wei:2019uqg}. However, the order parameter for a given system is often not unique and there are other possible choices for an order parameter \cite{NG_lecture}. In the free energy landscape of black hole phase transition, it is proposed that the black hole radius can be a proper order parameter to describe the black hole phase. In the example of Hawking-Page phase transition, the Schwarzschild Anti-de Sitter (SAdS) black hole phase has the non-zero black hole radius while the pure Anti-de Sitter (AdS) space phase has vanishing black hole radius \cite{Li:2020khm}. For the small/large RNAdS black hole phase transition, the three branches of black holes also have different black hole radii \cite{Li:2020nsy,Li:2021vdp}.

The kinetics of the black hole phase transition can be studied as follows. The evolution of the order parameter under the influence of thermal bath is assumed to be governed by the stochastic dynamics. In analogy to the random motion of particles suspended in thermal environment, there are three types of forces that determine the dynamics of the order parameter. The first force is the friction that results from the interaction between the black hole and the thermal bath. The second is the thermodynamic driving force, which is represented by the gradient of the generalized free energy function. The third is the stochastic force from the thermal environment. In this way, one can study the kinetics of black hole phase transition by using the Langevin equation for the dynamical trajectory evolution or the Fokker-Planck equation for the probabilistic evolution. The formalism has been applied to investigate the Markovian dynamics \cite{Wei:2020rcd,Li:2020spm,Wei:2021bwy,Cai:2021sag,Lan:2021crt,Li:2021zep,Yang:2021nwd,Mo:2021jff,Kumara:2021hlt,Li:2021tpu,Liu:2021lmr,Xu:2021usl,Du:2021cxs,Dai:2022mko} and the non-Markovian dynamics \cite{Li:2022ylz,Li:2022yti} of the black hole phase transitions.

In the present work, unlike the previous proposal \cite{Li:2021vdp} that the thermal bath is the effective description of the portion of the microscopic degrees of freedom of black hole, the thermal bath is considered as the real entity or the external environment that black hole can be in contact with. For the asymptotically flat cases, the radiations and the matter fields far away from the black hole horizon can be considered as the environment. For the AdS case, one can also imagine that there is a thermal bath at the AdS spatial infinity, just as the physical model studied in island formalism of Hawking radiation \cite{Penington:2019npb,Almheiri:2019psf,Almheiri:2019hni,Almheiri:2019yqk}. In this case, the AdS boundary does not behave like a mirror that reflects the radiation back into the black hole. It is transparent and allows the radiation or energy flow from the black hole to the bath or back \cite{Page:2015rxa}. The temperature of the environment or the thermal bath with which the system interacts is defined as the ensemble temperature. In this setup, the ensemble temperature is an external adjustable quantity, which is independent of the black hole parameters. In addition, due to the existence of the thermal environment, the fluctuating black holes will be generated from the local stable black holes by absorbing or radiating the matter or energy during the phase transition process.

\begin{figure}
  \centering
  \includegraphics[width=10cm]{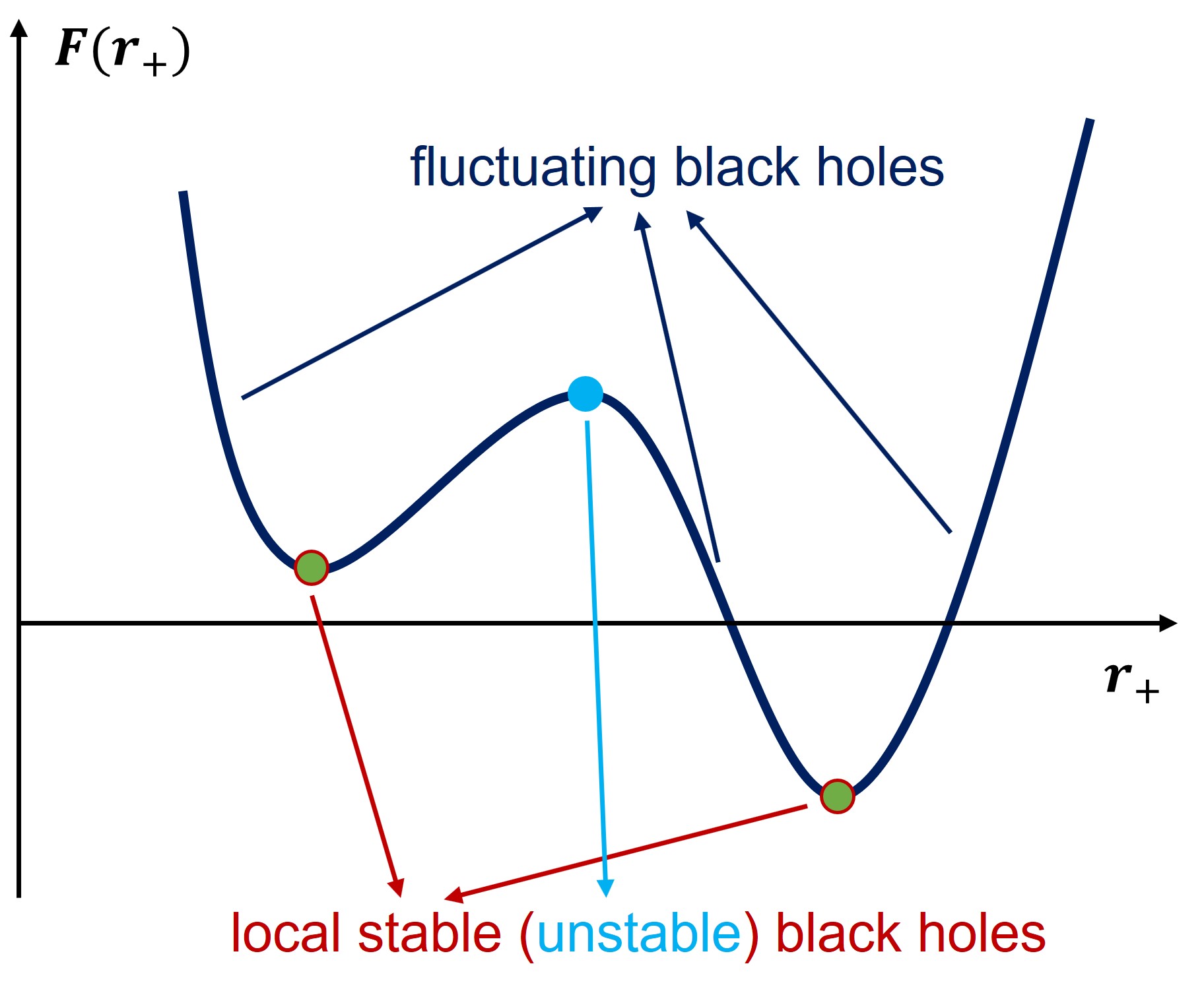}\\
  \caption{A sketch of the free energy landscape. The typical free energy landscape of the first order phase transition has the shape of double well when the generalized free energy $F(r_+)$ is plotted as the function of the order parameter $r_+$. On the landscape, the local minimum points (red) represents the local stable black holes, the local maximum point (light blue) represents the unstable black hole, and other points (dark blue) represents the fluctuating black holes.}
  \label{Free_Energy_Landscape}
\end{figure}

The free energy landscape provides an intuitive and quantitative description of the free energy topography, on which the local minimum/maximum points represents the local stable/unstable black holes while other points represents the fluctuating black holes \cite{Li:2020khm,Li:2020nsy,Li:2021vdp}. A sketch of the free energy landscape is presented in Figure \ref{Free_Energy_Landscape}. The fluctuating black holes act as the intermediate states during the phase transition process. It should be noted that the fluctuating black holes as well as the local stable and unstable black holes are distinguished by the order parameter (black hole radius) as discussed previously. Therefore, every point on the landscape, which represents a spacetime state in the phase transition process, should be described by a metric that solves the Einstein field equations. Then, the macroscopic thermodynamic quantities can be expressed as the functions of the order parameter (black hole radius). The main purpose of the current work is to compute the generalized free energy of the fluctuating black hole. It is expected that the generalized free energy should be related to the Euclidean gravitational action in the path integral approach to black hole thermodynamics.

In the Euclidean path integral approach to quantum gravity, the statistical ensemble consists of all the gravitational configurations that satisfy the specific boundary conditions. In our setup, the Euclidean gravitational configurations in the ensemble are specified by the following conditions: 1) the fixed ensemble temperature $T$; 2) the fixed horizon radius $r_+$ and the asymptotic AdS behavior on the external boundary. In principle, the partition function should take into account all the contributions from all the spacetime geometries that satisfy these conditions. It is well known that the saddle point approximation leads to the conventional semiclassical method to evaluate the partition function. In this way, one can consider only the contribution of the classical solution that solves Einstein field equations. More precisely, the partition function can be evaluated on the Euclidean fluctuating black hole with the fixed horizon radius $r_+$ and the fixed Euclidean time period $\beta$ related to the ensemble temperature $T$ as $\beta=1/T$. This type of classical geometry is known as the Euclidean black hole instanton with conical singularity at the horizon \cite{Susskind:1993ws,Banados:1993qp,Carlip:1993sa,Teitelboim:1994is}. Recall that the ensemble temperature is an external adjustable parameter, which is independent of the black hole parameters. Therefore, for the arbitrary period $\beta$ of the Euclidean time, the Euclidean manifold of the fluctuating black hole is not regular, but has a conical singularity. As is well known, there exists a special period (Hawking inverse temperature of the fluctuating black hole), for which the conical singularity disappears. However, for the arbitrary period, there is a two dimensional cone with non-zero deficit angle near the event horizon in the Euclidean manifold. Fortunately, the issue of how to deal with the conical singularity was previously investigated in \cite{Fursaev:1995ef,Mann:1996bi}. Thus, it is possible to calculate the partition function by evaluating the gravitational action on singular Euclidean gravitational instanton. The essential idea to derive the generalized free energy of the fluctuating black hole is to introduce an arbitrary fixed ensemble temperature. This approach was used to study the quantum corrections to the black hole entropy, which is called the off-shell method \cite{Solodukhin:1994yz,Fursaev:1994te,Solodukhin:1995ak,Frolov:1995xe,Frolov:1996hd,Bytsenko:1997ru,Solodukhin:2011gn}. We expect that the present work can provide a concrete and solid foundation for the free energy landscape formalism of black hole phase transition.

The rest of the paper is devoted to the computation of the generalized free energies of the SAdS black hole (Sec.\ref{Gen_free}), the RNAdS black hole (Sec.\ref{Gfe_RNAdS}), and the Kerr-AdS black hole (Sec.\ref{Gfe_KAdS}). The conclusion and discussion are presented in the last section.

\section{Generalized free energy landscape of Hawking-Page phase transition}
\label{Gen_free}

It is well known that Hawking-Page phase transition is the first order phase transition between the SAdS black hole and the thermal AdS space \cite{Hawking:1982dh}. In this section, our aim is to elucidate the idea of the calculation of the generalized free energy for the black hole phase transition.

To begin with, we recall that in the Gibbons-Hawking approach to black hole thermodynamics, the partition function of the canonical ensemble can be written in the form of the gravitational path integral \cite{Gibbons:1976ue}
\begin{eqnarray}\label{partition_func}
Z_{grav}(\beta)=\int D[g] e^{-I_E[g]}\simeq  e^{-I_E[g]} \;, 
\end{eqnarray}
where $I_E[g]$ is the Euclidean gravitational action and $\beta$ is the integral period of the Euclidean time. This functional is taken on all the Euclidean gravitational configurations that satisfy the given boundary conditions. For our purpose, the saddle point approximation is used.

The Euclidean Einstein-Hilbert action is given by \cite{Gibbons:1976ue}
\begin{eqnarray}\label{E_H_action}
I_E=-\frac{1}{16\pi} \int_{\mathcal{M}}\left( \mathcal{R} +\frac{6}{L^2}\right)\sqrt{g} d^4x\;,
\end{eqnarray}
where $\mathcal{R}$ is the Ricci scalar curvature. Note that there should be boundary terms that lead to the well defined variation for the equation of motion and the counter terms that cancel the divergence at the AdS spatial infinity. As shown in the following, we will employ the background subtraction trick to calculate the finite part of the Euclidean gravitational action \cite{Hawking:1982dh,Witten:1998qj}. The boundary terms will be cancelled in this procedure because the black hole correction to the AdS metric decays very rapidly at infinity. Thus, we only need to consider the bulk action.

The integral period $\beta$ of the Euclidean time in the partition function (\ref{partition_func}) is determined by the ensemble temperature $T$, which is given by 
\begin{eqnarray}
\beta=\frac{1}{T}\;.
\end{eqnarray}
As discussed in the introduction, the ensemble temperature is just the temperature of the thermal environment, which is an adjustable external parameter and independent of the black hole parameters in our setup.

To proceed, we consider the Euclidean SAdS black hole solution in four dimensions, which is described by the metric  
\begin{eqnarray}\label{SAdS}
ds^2=\left(1-\frac{2M}{r}+\frac{r^2}{L^2}\right)d\tau^2+
\left(1-\frac{2M}{r}+\frac{r^2}{L^2}\right)^{-1}dr^2+r^2d\Omega_2^2\;,
\end{eqnarray}
where $M$ is the mass of the SAdS black hole and $L$ is the AdS curvature radius. The mass of the SAdS black hole can be expressed as the function of the black hole radius $r_+$
\begin{eqnarray}
M&=&\frac{r_+}{2}\left(1+\frac{r_+^2}{L^2}\right) \;.\label{SAdS_mass}
\end{eqnarray}
As discussed in the introduction, black hole radius is treated as an order parameter. The fluctuating black hole with the given horizon radius $r_+$ has the mass given by Eq.(\ref{SAdS_mass}) and the corresponding geometry is then described by the metric Eq.(\ref{SAdS}). For example, consider the spacetime solution that has the order parameter $r_+=0$. From Eq.(\ref{SAdS_mass}), the mass of this spacetime solution is zero. The spacetime geometry is then described by the metric Eq.(\ref{SAdS}) with $M=0$. It is easy to see that this spacetime is just the pure AdS space. In this way, all the black hole states on the free energy landscape are described by this type of the SAdS metric towards the relation Eq.(\ref{SAdS_mass}).

\begin{figure}
  \centering
  \includegraphics[width=10cm]{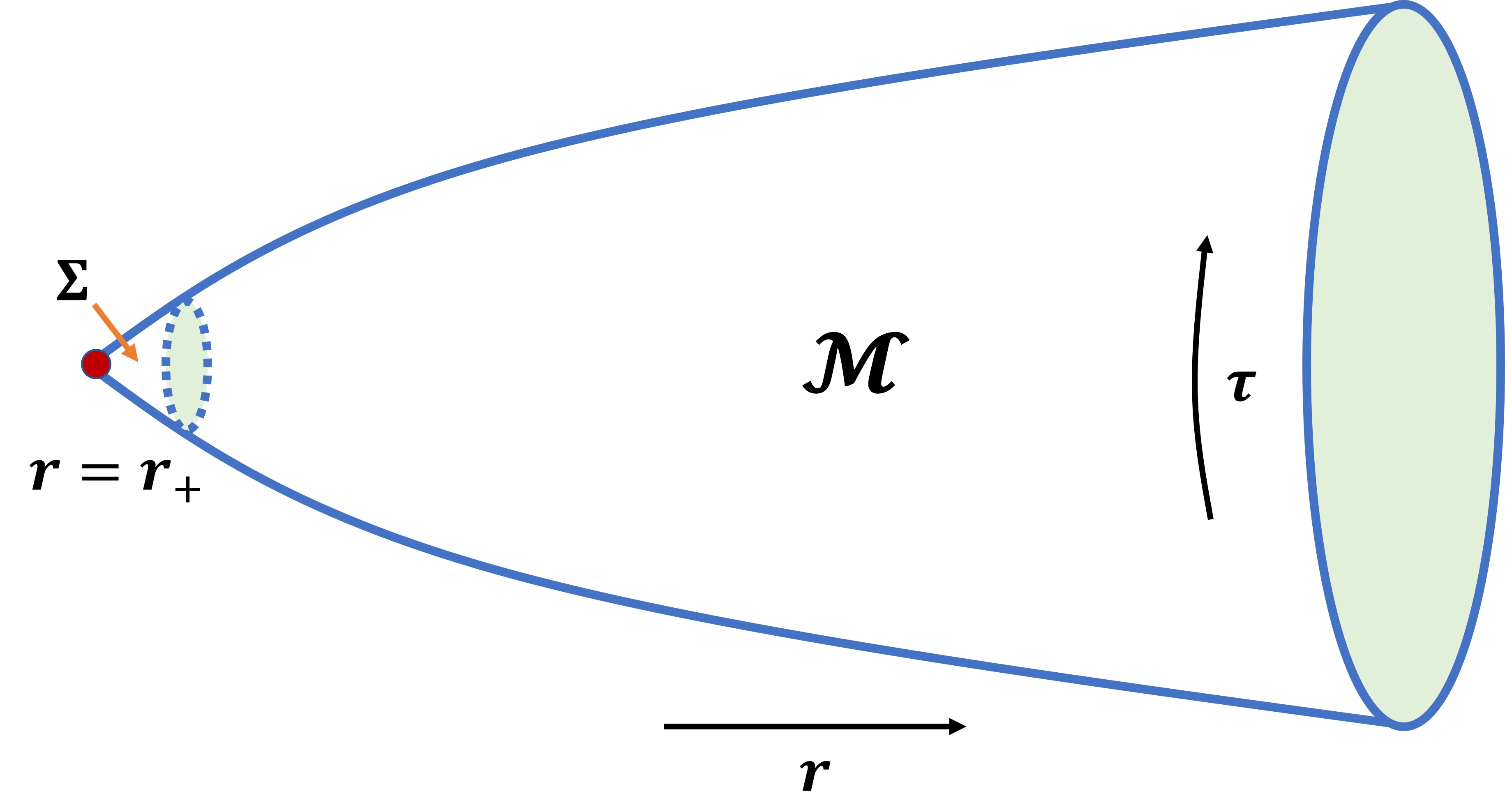}\\
  \caption{Euclidean geometry $\mathcal{M}$ of the fluctuating black hole with the event horizon radius $r_+$ and the arbitrary time period $\beta$. There is a conical singularity $\Sigma$ at the event horizon $r=r_+$. Every point in this two dimensional surface represents a sphere of radius $r$. When $\beta=\beta_H$, the singularity disappears and the corresponding Euclidean manifold is regular.}
  \label{Euclidean_Geometry}
\end{figure}

The presence of the environment/ensemble temperature $T=1/\beta$ implies that the periodicity of Euclidean time $\tau$ in the metric (\ref{SAdS}) is given by 
\begin{eqnarray}
0\leq \tau \leq \beta\;. 
\end{eqnarray}
In our setup, the period $\beta$ of the Euclidean time is independent of the black hole parameter $M$, or is irrelevant to the black hole radius $r_+$. By introducing the coordinate $\rho$ as
\begin{eqnarray}
d\rho=\frac{dr}{\sqrt{1-\frac{2M}{r}+\frac{r^2}{L^2}}}\;,
\end{eqnarray}
the near horizon metric of the Euclidean SAdS black hole can be approximated by 
\begin{eqnarray}\label{NH_metric}
ds^2\simeq \rho^2d\left(\frac{2\pi \tau}{\beta_H}\right)^2+d\rho^2+r_+^2d\Omega_2^2\;,
\end{eqnarray}
where $\beta_H$ is given by
\begin{eqnarray}
\beta_H=\frac{4\pi r_+}{\left(1+3r_+^2/L^2\right)}\;. 
\end{eqnarray}
For arbitrary period $\beta$ of the Euclidean time $\tau$, the metric (\ref{NH_metric}) represents the product manifold of a two dimensional cone and a two dimensional sphere. The Euclidean geometry of the fluctuating black hole is depicted in Figure \ref{Euclidean_Geometry}. The Euclidean manifold of the fluctuating black hole is not regular, but has a conical singularity \cite{Fursaev:1995ef,Mann:1996bi}. More precisely, there is a two dimensional cone with non-zero deficit angle $2\pi\left(1-\frac{\beta}{\beta_H}\right)$ near the event horizon $r=r_+$. It should be emphasized that the Euclidean SAdS metric (\ref{SAdS}) with the arbitrary time period $\beta$ satisfies the Einstein field equations except at the event horizon $r=r_+$. Therefore, to evaluate the partition function by using the semiclassical approximation method, one just needs to compute the Hilbert-Einstein action (\ref{E_H_action}) on the singular Euclidean gravitational instanton as depicted in Figure \ref{Euclidean_Geometry}.

It is well known that when the periodicity of the Euclidean time is taken to be the inverse Hawking temperature $\beta_H$, the near horizon metric represents a two dimensional disk rather than a singular cone. Then, the conical singularity disappears and the corresponding Euclidean manifold is regular. The physical implication of the condition $\beta=\beta_H$ for the regularity of the Euclidean manifold will be further explained at the end of this section.

Because the Euclidean gravitational instanton is singular, the gravitational bulk term contains the extra contribution from the conical singularity $\Sigma$ \cite{Fursaev:1995ef,Mann:1996bi}. It can be shown that this contribution is proportional to the horizon area times the deficit angle of the conical singularity \cite{Nishioka:2018khk,Gregory:2013hja,Burda:2015yfa}. It can be proved that \cite{Solodukhin:2011gn}
\begin{eqnarray}\label{can_R}
\int_{\mathcal{M}} \mathcal{R} =4\pi\left(1-\frac{\beta}{\beta_H}\right)\int_{\mathcal{H}} 1+\int_{\mathcal{M}/\Sigma} \mathcal{R}\;,
\end{eqnarray}
where $\mathcal{H}$ represents the event horizon $r=r_+$, and $\mathcal{M}/\Sigma$ represents the regular manifold by excising the conical singularity $\Sigma$.

For the regular part of the Euclidean gravitational instanton, the action becomes \begin{eqnarray}
I_{\mathcal{M}/\Sigma}=\frac{3}{8\pi L^2} \int_{\mathcal{M}/\Sigma}\sqrt{g} d^4x\;,
\end{eqnarray}
where we have used the fact that $\mathcal{R}=-\frac{12}{L^2}$ for the AdS solutions of the equations of motion. It is clear that this expression is divergent because the volume of the bulk is infinite. In order to regularize the bulk action, one can terminates the $r$ integral at a cutoff boundary $r=r_0$, subtract off the action of the pure AdS space, and finally take the limit of $r_0\rightarrow +\infty$ to obtain the finite part of the bulk action \cite{Hawking:1982dh,Witten:1998qj,Gibbons:2004ai}.

In this procedure, one has to match the SAdS metric with the background AdS metric at $r=r_0$. Thus the time coordinate $\tau_0$ of the AdS space is related to the time coordinate $\tau$ of the SAdS metric by the following condition 
\begin{eqnarray}
\left(1-\frac{2M}{r_0}+\frac{r_0^2}{L^2}\right) d\tau^2= \left(1+\frac{r_0^2}{L^2}\right) d\tau_0^2\;.
\end{eqnarray}
This in turn gives the time periods in the action integrals that are related by the relation 
\begin{eqnarray}
\beta_0=\beta\frac{\left(1-2M/r_0+r_0^2/L^2\right)^{1/2}}{\left(1+r_0^2/L^2\right)^{1/2}}
=\beta\left(1-\frac{M L^2}{r_0^3}+\mathcal{O}\left(\frac{1}{r_0^5}\right)\right)\;.
\end{eqnarray}
Thus, the bulk action of SAdS black hole with the subtraction of the background AdS action is then given by \cite{Gibbons:2004ai}
\begin{eqnarray}
I_{\mathcal{M}/\Sigma}=\frac{3}{8\pi L^2}\left(\beta\int_{r_+}^{r_0}\sqrt{g}drd\theta d\phi -\beta_0\int_{0}^{r_0}\sqrt{g_0}drd\theta d\phi \right)\;,
\end{eqnarray}
where $g_0$ is the Euclidean metric determinant of the pure AdS space. Note that there is a subtlety in the lower bound of the $r$ direction integral over the pure AdS space. However, for the even dimensions, the lower bound of the radial integration is just zero \cite{Gibbons:2004ai}.

At last, considering all the contributions discussed above, we have
\begin{eqnarray}
I_{E}&=&-\left(1-\frac{\beta}{\beta_H}\right)\frac{\mathcal{A}}{4}+  \frac{\beta}{2L^2}\left(r_0^3-r_+^3\right)- \frac{\beta_0}{2L^2} r_0^3 \;,\nonumber\\
&=&-\pi r_+^2\left(1-\frac{\beta}{\beta_H}\right)  +\frac{\beta M}{2}-\frac{\beta r_+^3}{2L^2}\nonumber\\
&=& \frac{\beta r_+}{2}\left(1+\frac{r_+^2}{L^2}\right)-\pi  r_+^2\;,
\end{eqnarray}
where $\mathcal{A}=4\pi r_+^2$ is the horizon area of the SAdS black hole and the cutoff surface $r_0$ is sent to infinity in the final result.

The free energy is defined as \cite{Landau:book}
\begin{eqnarray}\label{F_SAdS}
F&=&-\frac{1}{\beta} \ln Z_{grav}(\beta) \nonumber\\
&=&\frac{I_E}{\beta} \nonumber\\
&=&
\frac{r_+}{2}\left(1+\frac{r_+^2}{L^2}\right)-\pi T r_+^2\;.
\end{eqnarray}
This is just the generalized free energy of the fluctuating black hole previously defined from the thermodynamic relation $F=M-TS$ \cite{Li:2020khm}. To make it more explicit, we recall that the energy $E$ and entropy $S$ of a canonical ensemble at the temperature $T=1/\beta$ can be derived from the free energy as \cite{Landau:book}
\begin{eqnarray}\label{E_S}
E&=&\frac{\partial}{\partial \beta}\left(\beta F\right)=\frac{r_+}{2}\left(1+\frac{r_+^2}{L^2}\right)\;,\\
S&=&\beta\left(E-F\right)=\pi r_+^2\;.
\end{eqnarray}
It can be seen that the energy $E$ and the entropy $S$ are independent of the inverse temperature $\beta$. By identifying the energy $E$ as the black hole mass $M$, the thermodynamic definition of the generalized free energy is then given by 
\begin{eqnarray}\label{F_MTS}
F=M-TS=\frac{r_+}{2}\left(1+\frac{r_+^2}{L^2}\right)-\pi T r_+^2\;,
\end{eqnarray}
which coincides with the result Eq.(\ref{F_SAdS}) calculated from the gravitational action on the singular Euclidean instanton.

As shown, we have derived the generalized free energy of the fluctuating SAdS black hole from the gravitational action on the singular Euclidean instanton by using the path integral approach. In the equations (\ref{F_SAdS})-(\ref{F_MTS}), the black hole radius $r_+$ should be treated as the order parameter. In general, the order parameter can be interpreted as the one emergent from the underlying microscopic degrees of freedom of the black hole. For the SAdS black holes, the order parameter $r_+$ changes continuously from zero to infinity without any constraint. The generalized free energy $F$ is then the continuous function of the order parameter $r_+$ and the ensemble temperature $T$. This forms the free energy landscape for the SAdS black holes at the temperature $T$.

The second law of thermodynamics for an ensemble of arbitrary systems in contact with identical thermal baths is equivalent to the law that the free energy of the system can never increase. This can be stated as the minimum principle for the generalized free energy \cite{Landau:book}. By using the extremum condition $\partial F/\partial r_+=0$, one can obtain the condition for the local stable state in the thermodynamics as  
\begin{eqnarray}\label{Haw_temp}
T=\frac{1}{4\pi r_+}\left(1+\frac{3r_+^2}{L^2}\right)\;.
\end{eqnarray}
This is just the expression of Hawking temperature $T_H$, which implies that the Hawking temperature $T_H$ of the local stable black hole is equal to the ensemble's temperature $T$. Recall that this is also the condition that guarantees the regularity of the Euclidean SAdS black hole. Therefore, the local stability of the black hole is equivalent to the regularity of the corresponding Euclidean geometry. In fact, the condition (\ref{Haw_temp}) gives only the local extreme points on the landscape. The black hole being local stable or unstable state should be further determined by whether the secondary order derivative of the generalized free energy with respect to the order parameter is positive or negative.

Analogous to the ordinary thermodynamic system where the equation of state can be derived from the minimum principle of the generalized free energy \cite{Thorne:book}, the local stable condition (\ref{Haw_temp}) should be regarded as the equation of state for the black hole system. In this sense, the generalized free energy (\ref{F_SAdS}) or (\ref{F_MTS}) should be regarded as the off-shell free energy function at the arbitrary temperature $T$. By substituting Eq.(\ref{Haw_temp}) into Eq.(\ref{F_SAdS}) or Eq.(\ref{F_MTS}), one can obtain the on-shell value of the free energy for the SAdS black hole \cite{Hawking:1982dh}. Here, the on-shell or off-shell refers to whether the ensemble temperature is equal to the Hawking temperature or not.

The free energy landscape provides a pictorial and quantitative description of the generalized free energy topography. One can refer to Ref.\cite{Li:2020khm} for the detailed discussion on the free energy landscape. The minimum principle of the generalized free energy indicates that the local stable black holes are the extreme points on the free energy landscape. The thermodynamic stability of the local stable black hole and the kinetics of black hole phase transition were also discussed based on the free energy topography in \cite{Li:2020khm}.

\section{Generalized free energy landscape of RNAdS black holes}
\label{Gfe_RNAdS}

In this section, we derive the generalized free energy landscape of the the small/large RNAdS black hole phase transition in extended phase space \cite{Kubiznak:2012wp,Kubiznak:2014zwa,Kubiznak:2016qmn}. As shown in the last section, to evaluate the partition function in the semiclassical approximation, one just needs to compute the gravitational action of the Euclidean gravitational instanton with the conical singularity.

We start with the Euclidean metric of RNAdS black hole in four dimensions 
\begin{eqnarray}\label{RNAdS_metric}
ds^2=\left(1-\frac{2M}{r}+\frac{Q^2}{r^2}+\frac{r^2}{L^2}\right)d\tau^2+
\left(1-\frac{2M}{r}+\frac{Q^2}{r^2}+\frac{r^2}{L^2}\right)^{-1}dr^2+r^2d\Omega_2^2\;,
\end{eqnarray}
where $M$ is the mass, $Q$ is the electric charge, and $L$ is the AdS curvature radius. The mass of the fluctuating RNAdS black hole as the function of the black hole radius can be expressed as 
\begin{eqnarray}
M=\frac{r_+}{2}\left(1+\frac{r_+^2}{L^2}+\frac{Q^2}{r_+^2}\right)\;.
\end{eqnarray}
Note that the Euclidean time has the period $\beta$ determined by the ensemble temperature $T$. Therefore, the metric (\ref{RNAdS_metric}) with the general time period also describes the Euclidean gravitational instanton with the conical singularity, the geometry of which is depicted in Figure \ref{Euclidean_Geometry}. Analogous to the case of Hawking-Page phase transition, one can derive the generalized free energy of the fluctuating RNAdS black hole from the gravitational action of the singular Euclidean instanton by using the path integral approach.

The finite part of the Einstein-Hilbert action includes the contributions from the conical singularity and the AdS bulk. The singularity contribution is proportional to the horizon area of the RNAdS black hole times the deficit angle $\left(1-\frac{\beta}{\beta_H}\right)$, where $\beta_H$ is the inverse Hawking temperature of the RNAdS black hole:
\begin{eqnarray}
\beta_H=\frac{4\pi r_+}{1+3r_+^2/L^2-Q^2/r_+^2}\;.
\end{eqnarray} 
For the bulk contribution, we employ the background subtraction trick to compute the finite part. The matching of the RNAdS metric with the background AdS metric on the cutoff surface $r=r_0$ gives the following relation of the Euclidean time periods as
\begin{eqnarray}
\beta_0=\beta\frac{\left(1-2M/r_0+Q^2/r_0^2+r_0^2/L^2\right)^{1/2}}{\left(1+r_0^2/L^2\right)^{1/2}}
=\beta\left(1-\frac{M L^2}{r_0^3}+\mathcal{O}\left(\frac{1}{r_0^4}\right)\right)\;.
\end{eqnarray} 
It is easy to see that the large $r_0$ behavior for the RNAdS case is the same as that for the SAdS case. Therefore, the finite part of the Einstein-Hilbert action for the RNAdS bulk spacetime is the same as that for the SAdS case. However, One should also consider the electromagnetic field contribution, which is given by 
\begin{eqnarray}
I_{EM}=\frac{1}{16\pi} \int_{\mathcal{M}}F_{ab}F^{ab}\sqrt{g} d^4x=\frac{\beta Q^2}{2r_+}\;.
\end{eqnarray}
Thus the total Euclidean action for the fluctuating RNAdS black hole is given by 
\begin{eqnarray}
I_{E}&=&-\left(1-\frac{\beta}{\beta_H}\right)\frac{\mathcal{A}}{4}+  \frac{\beta}{2L^2}\left(r_0^3-r_+^3\right)- \frac{\beta_0}{2L^2} r_0^3
+\frac{\beta Q^2}{2r_+}\;,\nonumber\\
&=& \frac{\beta r_+}{2}\left(1+\frac{r_+^2}{L^2}+\frac{Q^2}{r_+^2}\right)-\pi  r_+^2\;,
\end{eqnarray}
where $\mathcal{A}=4\pi r_+^2$ is the event horizon area the RNAdS black hole and the cutoff boundary is sent to infinity in the last step.

The generalized free energy of the fluctuating RNAdS black hole is then given by \begin{eqnarray}\label{GibbsEq}
F=\frac{I_E}{\beta}=\frac{r_+}{2}\left(1+\frac{8}{3}\pi P r_+^2+\frac{Q^2}{r_+^2} \right)-\pi T r_+^2\;,
\end{eqnarray}
where $T$ is the ensemble temperature, and the effective thermodynamic pressure $P=\frac{3}{8\pi}\frac{1}{L^2}$ is introduced, with $L$ being the AdS curvature radius \cite{Kastor:2009wy,Dolan:2011xt}. From the generalized free energy, one can obtain the energy and the entropy of the fluctuating RNAdS black hole as 
\begin{eqnarray}
E&=&\frac{\partial}{\partial\beta}\left(\beta F\right)=\frac{r_+}{2}\left(1+\frac{8}{3}\pi P r_+^2+\frac{Q^2}{r_+^2}\right)\;,\\
S&=&\beta\left(E-F\right)=\pi r_+^2\;. 
\end{eqnarray}
It can be seen that the energy and the entropy of the fluctuating RNAdS black hole are also independent of the ensemble temperature $T$. Once again, the energy $E$ is identified as the black hole mass $M$ and the generalized free energy of the fluctuating RNAdS black hole obtained from the Euclidean path integral of gravitational action is also consistent with the thermodynamic definition $F=M-TS$ \cite{Li:2020nsy}.

By treating the black hole radius $r_+$ as the independent argument, one can formulate the generalized free energy landscape of the RNAdS black hole at the temperature $T$. The free energy landscape of the RNAdS black holes has the shape of double well \cite{Li:2020nsy}. The local stable state condition $\partial F/\partial r_+=0$ leads to three extreme points, i.e. three branches of the RNAdS black holes. They correspond to the small, the intermediate, and the large RNAdS black holes. In \cite{Li:2020nsy}, the stability of the three branches of RNAdS black holes was discussed based on the free energy landscape in detail. In addition, as observed in the case of Hawking-Page phase transition, the local stable state condition will lead to the condition that the ensemble temperature is equal to the Hawking temperature of the RNAdS black hole 
\begin{eqnarray}\label{RNAdS_Temp}
T=T_H=\frac{1}{4\pi r_+}\left(1+8\pi P r_+^2- \frac{Q^2}{r_+^2}\right)\;.
\end{eqnarray}
When this condition is satisfied, the Euclidean geometry of the RNAdS black hole is free of the conical singularity. In addition, the equation (\ref{RNAdS_Temp}) should be treated as the equation of state for the RNAdS black hole \cite{Kubiznak:2012wp}
\begin{eqnarray}
P=\frac{T}{2r_+}-\frac{1}{8\pi r_+^2}+\frac{Q^2}{8\pi r_+^4}\;.
\end{eqnarray}
For the arbitrary ensemble temperature, the generalized free energy (\ref{GibbsEq}) is off-shell. For the ensemble temperature satisfying the equation of state, the generalized free energy is on-shell. If substituting the equation (\ref{RNAdS_Temp}) into the expression Eq.(\ref{GibbsEq}) of the generalized free energy, one can obtain the on-shell value of the free energy for the RNAdS black hole \cite{Chamblin:1999hg,Caldarelli:1999xj}.

At last, we have to point out that the black hole radius $r_+$ as the order parameter of the small/large RNAdS black hole phase transition has a constraint from the non-negativity of the Hawking temperature $T_H$. The constraint on the order parameter is then given by 
\begin{eqnarray}
r_+\geq \frac{L}{\sqrt{6}}\left(\sqrt{1+\frac{12Q^2}{L^2}}-1\right)^{1/2}\;. 
\end{eqnarray}
This constraint implies that the RNAdS black hole cannot have arbitrary small event horizon due to the presence of electric charge \cite{Liu:2021lmr}.

\section{Generalized free energy landscape of Kerr-AdS black holes}
\label{Gfe_KAdS}

In this section, we derive the generalized free energy landscape for the rotating black hole in AdS space. Kerr-AdS black hole is a rotating black hole solution to Einstein equations in AdS space \cite{Carter:1968ks,Gibbons:2004uw}. In analogy to the RNAdS black holes, it was shown that the Kerr-AdS black holes in extended phase space also exhibit the Van der Waals-type liquid–gas phase transition \cite{Gunasekaran:2012dq,Altamirano:2013ane,Altamirano:2013uqa,Wei:2015ana}.

The Kerr-AdS black hole in four dimensions is described by the metric \cite{Carter:1968ks,Gibbons:2004uw}
\begin{eqnarray}
ds^2=-\frac{\Delta}{\rho^2}\left[dt-\frac{a}{\Xi}\sin^2\theta d\phi\right]^2
+\frac{\rho^2}{\Delta}dr^2+\frac{\rho^2}{\Delta_{\theta}}d\theta^2
+\frac{\Delta_{\theta}\sin^2\theta}{\rho^2}\left[adt-\frac{r^2+a^2}{\Xi}d\phi\right]^2\;,
\end{eqnarray}
where 
\begin{eqnarray}
\Delta&=&\left(r^2+a^2\right)\left(1+\frac{r^2}{L^2}\right)-2mr\;,\nonumber\\
\rho^2&=&r^2+a^2\cos^2\theta\;,\nonumber\\
\Delta_{\theta}&=&1-\frac{a^2}{L^2}\cos^2\theta\;,\nonumber\\
\Xi&=&1-\frac{a^2}{L^2}\;.
\end{eqnarray}

The physical mass $M$ and angular momentum $J$ of the Kerr-AdS black hole are related to the parameters $m$ and $a$ appearing in the metric by the following relations \cite{Gibbons:2004ai,Caldarelli:1999xj,Henneaux:1985tv}
\begin{eqnarray}
M=\frac{m}{\Xi^2}\;,\;\;\;J=\frac{ma}{\Xi^2}\;. 
\end{eqnarray}
Thus the mass and the angular momentum of the Kerr-AdS black hole can also be expressed as the functions of the horizon radius $r_+$: 
\begin{eqnarray}
M=\frac{r_+}{2\Xi^2}\left(1+\frac{a^2}{r_+^2}\right)\left(1+\frac{r_+^2}{L^2}\right)\;,\\
J=\frac{a r_+}{2\Xi^2}\left(1+\frac{a^2}{r_+^2}\right)\left(1+\frac{r_+^2}{L^2}\right)\;.
\end{eqnarray}
The other thermodynamic quantities are given by 
\begin{eqnarray}
T_H&=&\frac{r_+\left(1+a^2/L^2+3r_+^2/L^2-a^2/r_+^2\right)}{4\pi \left(r_+^2+a^2 \right)}\;,\\
S&=&\frac{\pi \left(r_+^2+a^2 \right)}{\Xi}\;,\\
\Omega_H&=&\frac{a
\left(1+r_+^2/L^2\right)}{r_+^2+a^2}\;.
\end{eqnarray}

The black hole radius $r_+$ as the order parameter of the Kerr-AdS black hole has to guarantee the non-negativity of the Hawking temperature $T_H$, which gives us the constraint on the order parameter $r_+$ as 
\begin{eqnarray}
r_+\geq \frac{L}{\sqrt{6}}\left[\sqrt{\left(1+\frac{a^2}{L^2}\right)^2+\frac{12a^2}{L^2}}-\left(1+\frac{a^2}{L^2}\right)\right]^{1/2}\;. 
\end{eqnarray}
This constraint condition implies that the Kerr-AdS black hole cannot have arbitrary small event horizon due to the presence of the rotations \cite{Yang:2021nwd}.

We now proceed to derive the gravitational action on the Euclidean Kerr-AdS geometry with the arbitrary time period $\beta$. We firstly consider the contribution of the conical singularity to the Euclidean action. For the rotating black hole, the geometry is more complex because near the conical singularity the Euclidean geometry is no longer the product of the horizon sphere and the two dimensional cone as the static black hole case \cite{Mann:1996bi}. Instead, the near horizon geometry of the stationary rotating black hole is a nontrivial foliation of the horizon surface, which is shown to share certain common features with the static case \cite{Mann:1996bi}. It is further argued that the curvature singularity at the horizon of a stationary rotating black hole behaves in the same way as in the static case. This is to say that Eq.(\ref{can_R}) is still valid in the stationary rotating case \cite{Solodukhin:2011gn}. Thus, the contribution to the action from the conical singularity is given by 
\begin{eqnarray}
I_C=-\left(1-\frac{\beta}{\beta_H}\right)\frac{\mathcal{A}}{4}\;,
\end{eqnarray}
where $\mathcal{A}=\frac{4\pi \left(r_+^2+a^2 \right)}{\Xi}$ is the horizon area of the Kerr-AdS black hole.

We then compute the gravitational action on the regular bulk part of the Euclidean Kerr-AdS geometry. In order to subtract the volume contribution of the background AdS space, one has to match the geometry of the Kerr-AdS metric with the background AdS at the cutoff surface $r=r_0$, which leads to the matching condition as \cite{Gibbons:2004ai} 
\begin{eqnarray}
\Delta\left(r_0\right) d\tau^2= \Delta_0\left(r_0\right) d\tau_0^2\;,
\end{eqnarray}
where $\Delta_0$ denotes $\Delta(m=0)$. Note that the metric for $m=0$ is just the metric of the pure AdS space in non-standard “spheroidal” coordinates. For the large cutoff $r_0$, the Euclidean action integral over the background AdS space has the time period 
\begin{eqnarray}
\beta_0=\beta \left[\frac{\left(r_0^2+a^2\right)\left(1+\frac{r_0^2}{L^2}\right)-2mr_0}{\left(r_0^2+a^2\right)\left(1+\frac{r_0^2}{L^2}\right)}\right]^{1/2}= \beta \left(1-\frac{m L^2}{r_0^3}+\mathcal{O}\left(\frac{1}{r_0^5}\right)\right)\;.
\end{eqnarray}

The finite part of the bulk gravitational action is then obtained as the difference between the Kerr-AdS black hole and the background AdS space. It is given by 
\begin{eqnarray}
I_B=\frac{3}{8\pi L^2}\left(\beta\int_{r_+}^{r_0}\sqrt{g}drd\theta d\phi -\beta_0\int_{0}^{r_0}\sqrt{g_0}drd\theta d\phi \right)\;,
\end{eqnarray}
where $g$ and $g_0$ are the Euclidean metric determinants of the Kerr-AdS black hole and the background AdS space, respectively. It is easy to see that $g=g_0=\frac{\left(r^2+a^2\cos^2\theta\right)}{\Xi}\sin\theta$ in the $\left(\tau,r,\theta,\phi\right)$ coordinates. Performing the integral and taking the limit $r_0\rightarrow +\infty$, one can finally obtain 
\begin{eqnarray}
I_B=\frac{\beta}{2\Xi}\left[m-\frac{r_+\left(r_+^2+a^2\right)}{L^2}\right]\;.
\end{eqnarray}

Adding up the contributions from the conical singularity and the AdS bulk, one can obtain the gravitational action of the singular Euclidean instanton in a closed form as
\begin{eqnarray}
I_E&=&-\left(1-\frac{\beta}{\beta_H}\right)\frac{\mathcal{A}}{4}+\frac{\beta}{2\Xi}\left[m-\frac{r_+\left(r_+^2+a^2\right)}{L^2}\right]\;,\nonumber\\
&=&\frac{\beta r_+}{2\Xi}\left(1+\frac{r_+^2}{L^2}\right)-\frac{\pi \left(r_+^2+a^2 \right)}{\Xi}\;.
\end{eqnarray}
For the conical ensemble with the angular momentum, the generalized free energy is defined as 
\begin{eqnarray}
F&=&\frac{I_E}{\beta}+\Omega_H J\nonumber\\
&=&\frac{r_+}{2\Xi^2}\left(1+\frac{a^2}{r_+^2}\right)\left(1+\frac{r_+^2}{L^2}\right)-\frac{\pi T \left(r_+^2+a^2 \right)}{\Xi}\;, 
\end{eqnarray}
which is obviously consistent with the thermodynamic definition $F=M-TS$ in \cite{Yang:2021nwd}. The generalized free energy of the Kerr-AdS black hole as the function of black hole radius $r_+$ can be shown to have the shape of double well \cite{Yang:2021nwd}, which forms the free energy landscape of the Kerr-AdS black hole at the temperature $T$.

The equilibrium state condition $\partial F/\partial r_+=0$, which is equivalent to $T=T_H$, leads to three extreme points. They correspond to the small, the intermediate, and the large Kerr-AdS black holes. In \cite{Yang:2021nwd}, the stability of the three branches and the kinetics of phase transition of the Kerr-AdS black holes were discussed based on the free energy landscape in detail.

\section{Conclusion and discussion}
\label{Conc}

In summary, by employing the Gibbons-Hawking path integral approach to black hole thermodynamics, we show that the generalized free energy of the fluctuating black hole generated during the phase transition process can be derived from the partition function of the canonical ensemble evaluated on the Euclidean gravitational instanton with the conical singularity in the semiclassical approximation. To illustrate this point, we have presented the computation of the generalized free energies of the Schwarzschild-AdS black hole, the RNAdS black hole, and the Kerr-AdS black hole explicitly. We expect that the current work can provide a concrete and solid foundation for the free energy landscape formalism of the black hole phase transition.

Let us make some more discussion on the meaning of the generalized free energy. One can regard the fluctuating black holes with the Hawking temperature not equal to the ensemble temperature as being in the non-equilibrium states with the thermal baths. In this case, the fluctuating black holes will relax to the local stable black holes. The generalized free energy can be considered as the thermodynamic potential of the non-equilibrium states \cite{Thorne:book,Malakhov:1994}. If ignoring the fluctuating force (noise) of the thermal bath, the relaxation process from the initial non-equilibrium black hole with $T\neq T_H$ to the final equilibrium black hole with $T=T_H$ is completely determined by the gradient force of the non-equilibrium potential.

\begin{figure}
  \centering
  \includegraphics[width=8cm]{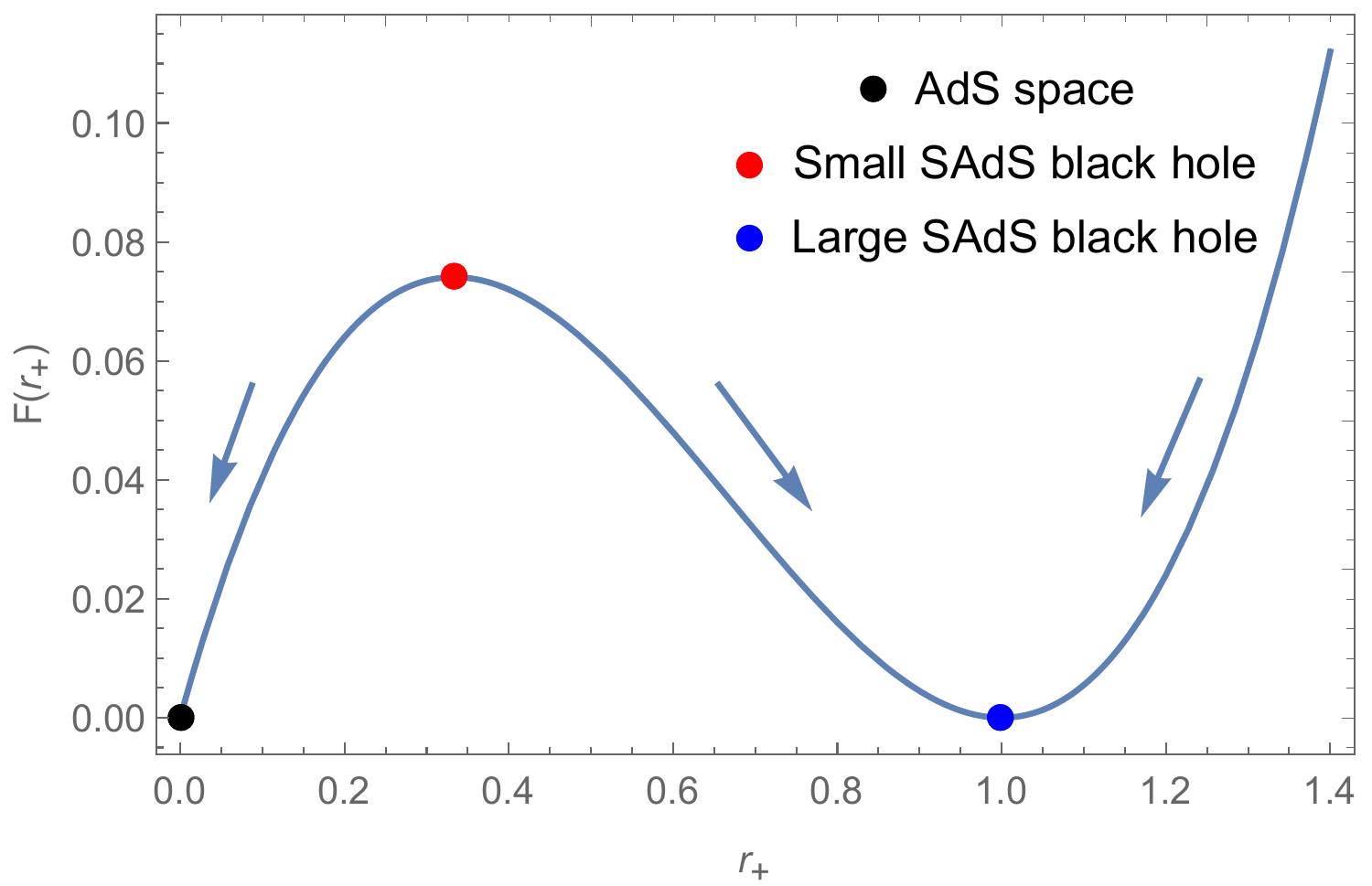}
    \includegraphics[width=8cm]{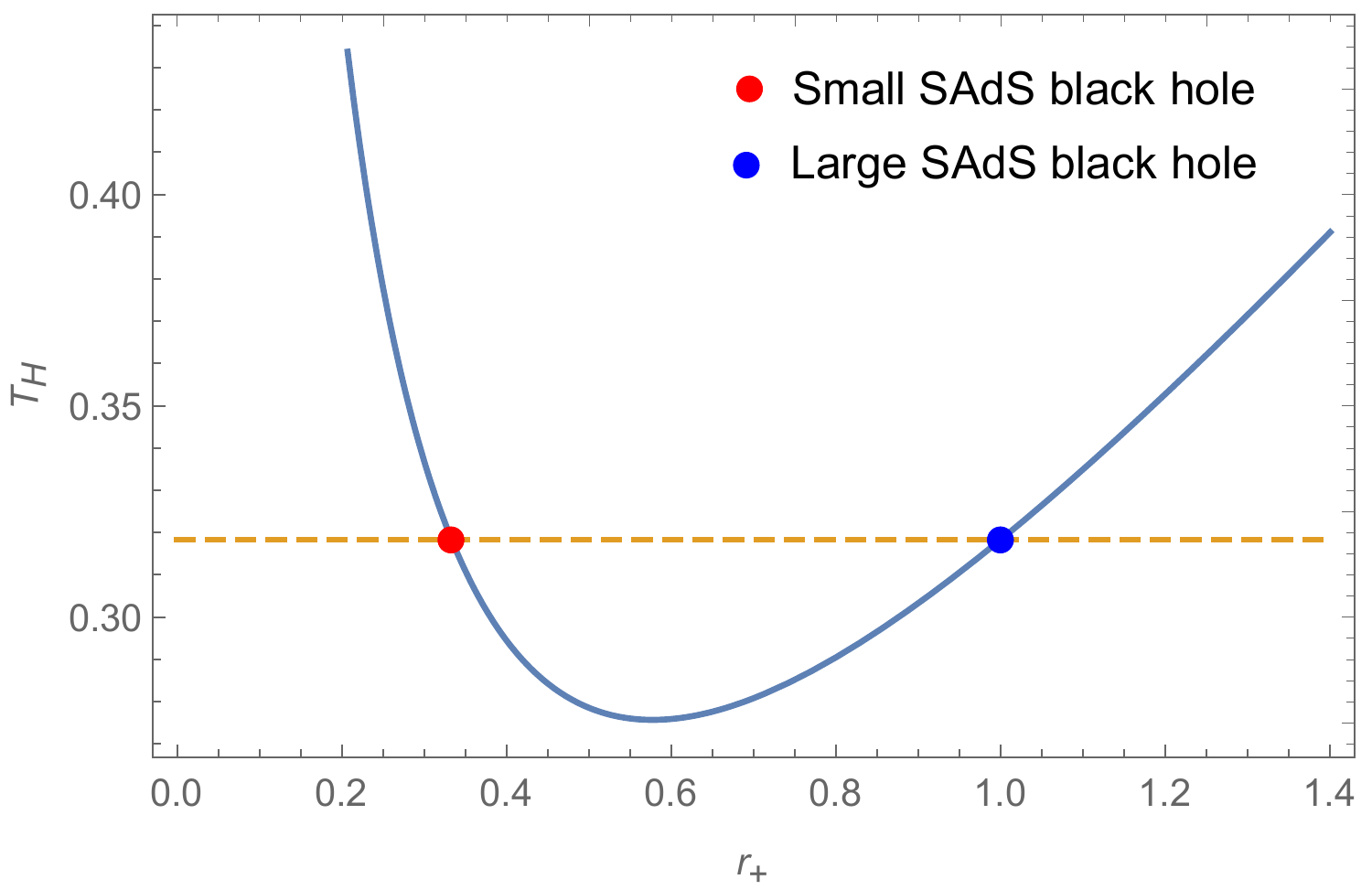}\\
  \caption{The plots of the generalized free energy (left panel) and the Hawking temperature (right panel) of the fluctuating SAdS black hole as the functions of the black hole radius. In the left panel, the arrows indicate the directions of the relaxation process for the non-equilibrium black holes without the fluctuating forces. In the right panel, the dashed line represents the ensemble temperature. Here, the ensemble temperature is selected to be the critical temperature of Hawking-Page phase transition. The AdS radius $L$ is set to unity. }
  \label{Free_energy_HP}
\end{figure}

To give a physical picture of the above statement, let us discuss the free energy landscape for the Hawking-Page phase transition. In Figure \ref{Free_energy_HP}, we plot the generalized free energy and the Hawking temperature of the fluctuating SAdS black hole. The black, the red, and the blue points represents the AdS space, the small SAdS black hole, and the large SAdS black hole, respectively. The ensemble temperature $T$ is selected to be the critical temperature of Hawking-Page phase transition $T_{HP}=1/\pi L$. For convenience, we denote the radius of the small/large SAdS black hole as $r_s/r_l$.

When $r_+<r_s$, $T_H>T$. Then the fluctuating black holes with the radius smaller than $r_s$ will radiate out energy to baths, which will reduce their masses and in turn decrease their horizon radii because the mass is the monotonic increasing function of the radius. Finally, these black holes will relax to the AdS space. When $r_s<r_+<r_l$, $T_H<T$. In this case, the fluctuating black holes will absorb energy or matter from baths, which will increase their masses and horizon radii. Finally, these black holes will relax to the large  SAdS black hole. When $r_+>r_l$, $T_H>T$. In this case, the fluctuating black holes will also relax to the large SAdS black hole. In the left panel of Figure \ref{Free_energy_HP}, we have explicitly indicated the the directions of the relaxation process for the non-equilibrium black holes without the fluctuating forces. The above analysis can also be performed for the RNAdS black hole and the Kerr-AdS black hole. Therefore, the free energy landscape gives a physical picture of the relaxation process of the non-equilibrium black holes without the fluctuating forces.

However, this is not adequate for studying the kinetics of the phase transition. If only considering the gradient force from the free energy landscape, the local stable SAdS black hole can never make a transition and switch to the AdS space, and the reverse statement is also true. It is the fluctuating force (thermal noise) from the bath that makes the transition process possible. Therefore, the dynamics of black hole phase transition is described by the Langevin equation for the stochastic evolution or equivalently the Fokker-Planck equation for the probabilistic evolution. In this sense, the late time stationary Boltzmann distribution $P\sim e^{-\beta F}$ for the fluctuating black holes can be obtained from the Fokker-Planck equation.

At last, we compare our assumptions made in the current work with those in the previous work on the free energy landscape of black hole phase transition. First, in the previous studies \cite{Li:2020khm,Li:2020nsy,Li:2021vdp}, it was stated that the fluctuating black holes are not the solutions to Einstein field equations. In this work, we clarify that the fluctuating black holes as well as the local stable/unstable black holes are all the solutions to Einstein field equations. This is to say that their geometries can be described by the metric that solves the Einstein equations. As we have discussed, order parameter (black hole radius) determines the black hole parameters (mass, angular momentum, et.al), and in turn black hole parameter determines the geometry of the fluctuating black hole. However, in evaluating the partition function at the arbitrary ensemble temperature by using the saddle point approximation, the Euclidean geometry of the fluctuating black hole is singular due to the conical singularity at the event horizon. As shown, the Euclidean geometry of the local stable black hole is regular because the Hawking temperature of the local stable black hole is equal to the ensemble temperature and the conical singularity is cancelled.

Second, in the previous work \cite{Li:2020khm,Li:2020nsy,Li:2021vdp}, it was assumed that there is no constraint on the order parameter (event horizon radius) of the fluctuating black hole. This was based on the statement that the fluctuating black holes are not the solutions to Einstein filed equations. If so, the Hawking temperature of the fluctuating black hole would not make sense any longer. The black hole radius is then not required to guarantee the non-negativity of the Hawking temperature of the fluctuating black hole. This is to say that the size of the fluctuating black hole can vary from zero to arbitrary large value. In the present work, we have assumed that the fluctuating black holes as well as the local stable/unstable black holes are all the solutions to Einstein field equations. In this sense, the relation between the Hawking temperature and the event horizon radius for the fluctuating black hole imposes the constraint on the order parameter (black hole radius). Thus, it seems not quite appropriate to assume that the order parameter of the fluctuating black hole can take arbitrary value. It is shown that for the Hawking-Page phase transition, the Hawking temperature of the SAdS black hole is always positive in spite of the event horizon radius. In this case, there is no constraint on the order parameter. For the RNAdS black holes and the Kerr-AdS black holes, the constraints on the order parameter can be properly carried out in order to guarantee the non-negativity of the Hawking temperature.

Third, in the previous work \cite{Li:2020khm,Li:2020nsy,Li:2021vdp}, it was assumed that the thermal bath stems from the effective description of the partial microscopic degrees of freedom of the black hole. In this work, we  point out that there are identical external thermal environments/baths with which each system in the ensemble can be in contact. For the AdS case considered in the present work, the thermal bath can be placed at the AdS spatial infinity and the AdS boundary is transparent to allow the energy or radiation to pass through. In studying the kinetics of black hole phase transition, the evolution of the order parameter is then governed by the gradient force from the free energy landscape as well as the friction force and the stochastic noise from the thermal baths. Thus, the friction coefficient reflects the interaction strength between the fluctuating black hole and the external thermal bath. Except this point, all the conclusions that obtained in the previous work \cite{Li:2020khm,Li:2020nsy,Li:2021vdp,Li:2022ylz,Li:2022yti} do not change.

\end{document}